\documentclass[aps,prl,twocolumn,showpacs,showkeys,groupaddresses,superscriptaddress,preprintnumbers]{revtex4-1}
\usepackage{amsmath,amssymb,bm,mathrsfs}
\usepackage{srcltx}
\usepackage{dsfont}
\usepackage{epsfig}
\usepackage{slashed}
\usepackage{bbold}
\usepackage{psfrag}
\usepackage{xcolor}
\PassOptionsToPackage{caption=false}{subfig}
\usepackage{subfig}
\usepackage{xfrac}
\usepackage{multirow}
\usepackage{booktabs}
\usepackage[colorlinks=true,linkcolor=blue,citecolor=blue,urlcolor=violet]{hyperref}

\def\ds{\displaystyle}
\newcommand{\be}{\begin{equation}}
\newcommand{\ee}{\end{equation}}
\newcommand{\bea}{\begin{eqnarray}}
\newcommand{\eea}{\end{eqnarray}}

\newcommand{\nn}{\nonumber}
\def\({\left(}
\def\){\right)}

\def\mw{m_{W}}

\def\W{{\rm{\sc{W}}}}
\def\Y{{\rm{\sc{Y}}}}

\usepackage{soul}

\begin{document}

\title{Energy helps accuracy:  electroweak precision tests at hadron colliders\vspace{0.4cm}} 

\author{Marco Farina}
\email{farina.phys@gmail.com}
\affiliation{
New High Energy Theory Center, Department of Physics, Rutgers University,  \\
136 Frelinghuisen Road, Piscataway, NJ 08854, USA
} 

\author{Giuliano Panico}
\email{gpanico@ifae.es}
\affiliation{IFAE, Universitat Aut\`onoma de Barcelona, E-08193 Bellaterra, Barcelona, Spain} 

\author{Duccio Pappadopulo}
\email{duccio.pappadopulo@gmail.com}
\affiliation{
Center for Cosmology and Particle Physics, \\
Department of Physics, New York University, New York, NY 10003, USA
}

\author{Joshua T. Ruderman}
\email{ruderman@nyu.edu}
\affiliation{
Center for Cosmology and Particle Physics, \\
Department of Physics, New York University, New York, NY 10003, USA
} 

\author{Riccardo Torre}
\email{riccardo.torre@cern.ch}
\affiliation{Institut de Th\'eorie des Ph\'enomenes Physiques, EPFL, Lausanne, Switzerland}

\author{Andrea Wulzer}
\email{andrea.wulzer@pd.infn.it}
\affiliation{Institut de Th\'eorie des Ph\'enomenes Physiques, EPFL, Lausanne, Switzerland}
\affiliation{Theoretical Physics Department, CERN, Geneva, Switzerland}
\affiliation{Dipartimento di Fisica e Astronomia, Universit\'a di Padova, Italy}

\begin{abstract}\vspace{0.4cm}
We show that high energy measurements of Drell-Yan at the LHC can serve as electroweak precision tests.  Dimension-6 operators, from the Standard Model Effective Field Theory, modify the high energy behavior of electroweak gauge boson propagators.  Existing measurements of the dilepton invariant mass spectrum, from neutral current Drell-Yan at 8 TeV, have comparable sensitivity to LEP\@.   We propose measuring the transverse mass spectrum of charged current Drell-Yan, which can surpass LEP already with 8 TeV data. The 13 TeV LHC will elevate electroweak tests to a new precision frontier. \vspace{0.8cm}
\end{abstract}

\preprint{CERN-TH-2016-205}
\maketitle
\noindent {\bf  Introduction.---} Hadron colliders are often viewed as ``discovery machines."  They have limited precision, due to their messy QCD environments, but their high Center of Mass (CoM) energies allow them to directly produce new, heavy, particles.  Hadron colliders are often contrasted with less energetic lepton colliders, which can reach high precision to indirectly probe new heavy physics, as exemplified by LEP, which tested the electroweak sector of the Standard Model (SM) with unprecedented per-mill accuracy~\cite{ALEPH:2005ab}. 

The flaws in this argument are well known to practitioners of Effective Field Theory (EFT).  
Probing heavy new physics, described by a mass scale $M$, at energies $E\ll M$, gives a correction to observables scaling as $(E/M)^{n}$, for some $n\geq0$.
For those observables with $n>0$, hadron colliders benefit from the high CoM~energy~\cite{Domenech:2012ai,Biekoetter:2014jwa,Azatov:2015oxa,Dror:2015nkp,Butter:2016cvz,Falkowski:2016cxu}. 
Is the energy enhancement at hadron colliders sufficient to beat the precision of lepton colliders?

We address this question within the SM EFT~\cite{Buchmuller:1985jz,Grzadkowski:2010es}.   We study the effect of ``universal'' new physics~\cite{Barbieri:2004qk,Wells:2015uba,Wells:2015cre} on neutral and charged Drell-Yan (DY)~\cite{Drell:1970wh} processes: $pp\to \ell^+\ell^-$ and $pp\to \ell\nu$. Universal theories include scenarios with new heavy vectors that mix with SM ones~\cite{Salvioni:2009mt,Contino:2010mh,delAguila:2010mx,deBlas:2012qp,Lizana:2013xla,Pappadopulo:2014qza}, new electroweak charged particles~\cite{Cirelli:2005uq}, and electroweak gauge boson compositeness~\cite{Liu:2016idz}.
The effects of universal new physics on the DY process can be parameterized as modifications of electroweak gauge boson propagators and encapsulated in the ``oblique parameters"~\cite{Peskin:1990zt}.  At leading order in a derivative expansion they correspond to $\hat{\textrm{S}}$, $\hat{\textrm{T}}$, $\W$, and $\Y$~\cite{Barbieri:2004qk}, which modify the $\gamma$, $Z$, and $W$ propagators.  The effects of $\hat{\textrm{S}}$ and $\hat{\textrm{T}}$ on DY processes do not grow with energy, making it difficult for the LHC to surpass stringent constraints from LEP~\cite{ALEPH:2005ab}.  On the other hand, $\W$ and $\Y$, which are generated by the dimension-6 operators of table~\ref{tab:op}, give rise to effects that grow with energy.

We find that neutral DY has comparable sensitivity to $\W$ and $\Y$ as LEP, already at 8\,TeV\@. This sensitivity follows from the growth in energy, as well as the percent-level precision achieved by LHC experiments~\cite{Chatrchyan:2011nv,CMS:2011aa,Aad:2011dm,Chatrchyan:2014mua,CMS:2014jea,Aad:2016naf,Aad:2016zzw}, Parton Distribution Function (PDF) determination, and NNLO calculations~\cite{Hamberg:1990np,Anastasiou:2003yy,Anastasiou:2003ds,Melnikov:2006kv,Catani:2009sm,Gavin:2010az,Li:2012wna}. We propose that the LHC can carry out similar measurements in charged DY (using the transverse mass spectrum), which with current data is sensitive to $\W$ far beyond LEP\@.   
We project the sensitivity of the 13 TeV LHC, and future hadron colliders, and find  spectacular  reach to probe $\W$ and $\Y$.\\
\indent While we propose to use DY for electroweak precision tests, previous studies have shown DY can probe 4-fermion contact 
operators~\cite{Cirigliano:2012ab,Aad:2012bsa,Chatrchyan:2012hda,Chatrchyan:2013lga,deBlas:2013qqa,Aad:2014wca,Khachatryan:2014tva,Khachatryan:2014fba}, 
 the running of electroweak gauge couplings~\cite{Rainwater:2007qa,Alves:2014cda},  and quantum effects from superpartners~\cite{Brensing:2007qm,Dittmaier:2009cr}.

\begin{table}[htb]
\renewcommand{\arraystretch}{1.6}
\small
\begin{tabular}{c|c|c}
& ~~universal  form factor~($\mathcal L$)~~ & ~~contact operator~($\mathcal L'$)~~ \\ \hline\hline
~W~ & $-\frac{\W}{4\mw^2}(D_\rho W_{\mu\nu}^a)^2$ & $-\frac{g_2^2 \W}{2\mw^2}{J_L}_{\mu}^a{J_L}_{a}^\mu$ \\ \hline
~Y~ &  $-\frac{\Y}{4\mw^2}(\partial_\rho B_{\mu\nu})^2$ & $-\frac{g_1^2 \Y}{2\mw^2}{J_Y}_{\mu}{J_Y}^\mu$ \\
\end{tabular} \caption{\label{tab:op}\it 
The parameters $\W$ and $\Y$  in their ``universal" form ({\it left}), and as products of currents related by the equation of motion ({\it right}). 
We dropped corrections to trilinear gauge couplings.}
\end{table}

\begin{figure*}[tb]
\centering
\includegraphics[width=0.85\linewidth]{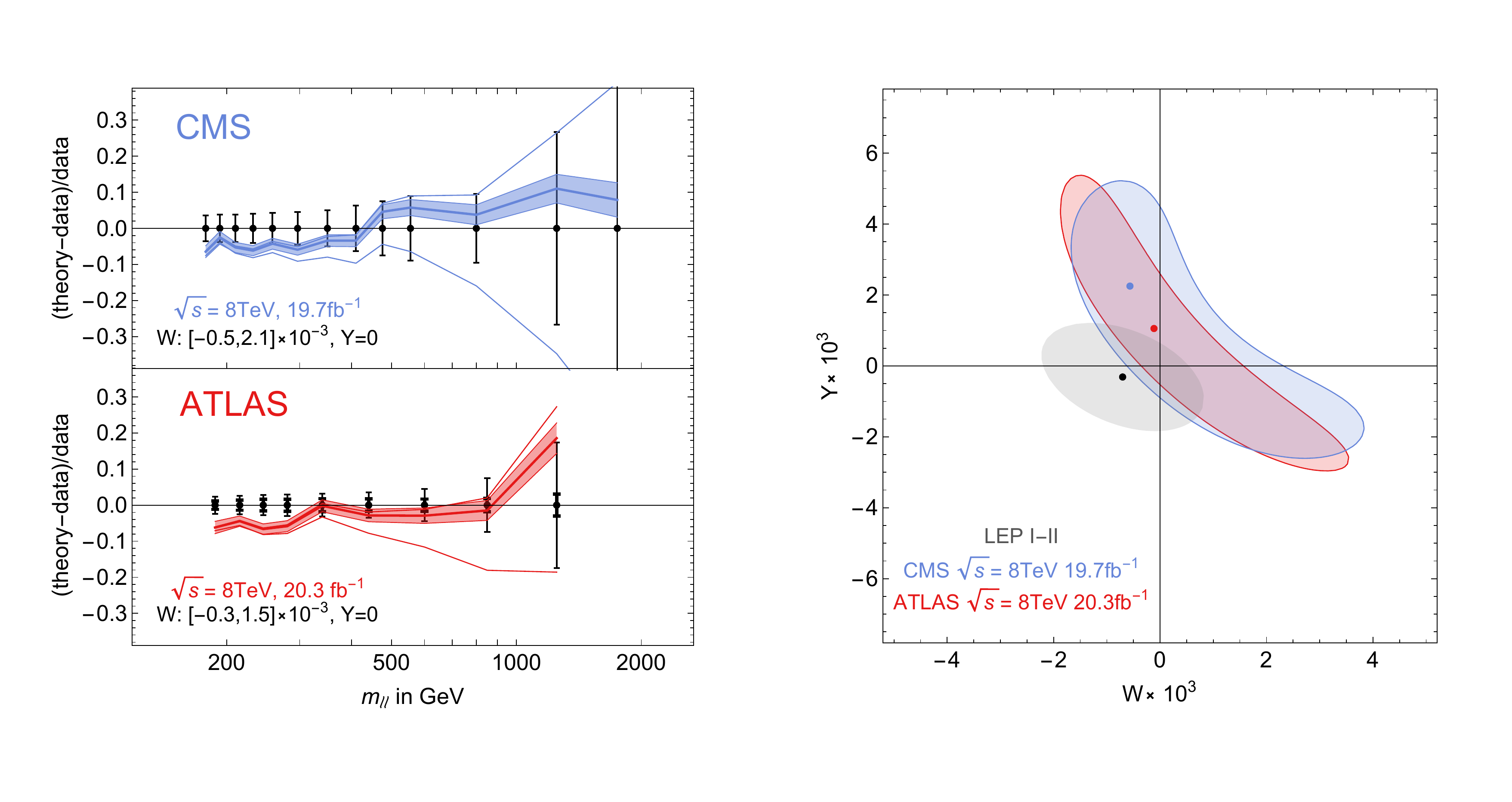}
\vspace{-.3cm}
\caption{
\label{fig1} \it Fit to CMS \cite{CMS:2014jea} and ATLAS \cite{Aad:2016zzw} dilepton invariant mass distributions measured at 8\,TeV.
 \textbf{Left:}~ comparison of data and SM prediction. The error bars include the fractional experimental uncertainties, while the thickness of the SM predictions include uncertainties from PDF and scale variation. The smaller error bars in the ATLAS plot show the systematic uncertainties.  We also show how the central value of the theoretical prediction changes when $\W$ varies within its 95\% CL range. \textbf{Right:}~ 95\%\,CL constraints in the \W-\Y ~plane.}
\label{fig:data}
\end{figure*}

\noindent {\bf  EWPT from DY.---}
The $4$ parameters $\hat{\textrm{S}}$, $\hat{\textrm{T}}$, $\W$, and $\Y$ modify the SM neutral ($\gamma$, $Z$) and charged ($W^\pm$) vector boson propagators as \footnote{These modified propagators encapsulate all new physics effects because they are written in the field basis where the vector boson interactions with fermions are identical to those of the SM, once expressed in terms of the input parameters $\alpha_{\rm{em}}$, $G_F$, and $m_Z$. This explains the mismatch with  Ref.~\cite{Barbieri:2004qk}, where a different basis is used.}
\bea\label{eq:propagators}
 P_{\hspace{-1pt}{N}}\hspace{-1pt}&=&\ds\hspace{-2.5pt}\left[\hspace{-2pt}
\begin{array}{cc}
 \frac{1}{q^2}-\frac{t^{2}\W+\Y}{m_Z^2} & \frac{t \left(\left(\Y+{\rm\hat{T}}\right) c^2+s^2 \W-{\rm\hat{S}}\right)}{\left(c^2-s^2\right) \left(q^2-m_Z^2\right)}+\frac{t (\Y-\W)}{m_Z^2} \\
 \star & \frac{1+ {\rm\hat{T}}-\W-t^{2}\Y}{q^2-m_Z^2}-\frac{t^{2}\Y+\W}{m_Z^2} \\
\end{array}
\hspace{-2pt}
\right]\nn\\[5pt]
P_{\hspace{-1pt}{C}}\hspace{-1pt}&=&
\begin{array}{c}
\frac{1+\left( \left({\rm\hat{T}}-\W-t^2 \Y\right)-2 t^2 \left({\rm\hat{S}}-\W-\Y\right)\right)/(1-t^2)}{(q^{2}-m_W^2)}-\frac{\W}{m_W^2}\,,
\end{array}
\eea
where $q$ is the four-momentum and $s$, $c$, and $t$ are the sine, cosine, and tangent of the Weinberg angle.  The parameters $\hat{\textrm{S}}$ and $\hat{\textrm{T}}$ have normalizations that differ from the conventional normalizations~\cite{Peskin:1990zt} as follows: $\hat S = \alpha/(4s^2) \, S$ and $\hat T = \alpha \,T$.  All $4$ parameters are constrained at the few per mille level, mainly from precision data collected at LEP \cite{Falkowski:2015krw} and from $W$ boson mass measurements at the Tevatron~\cite{Aaltonen:2012bp,Abazov:2012bv}.

In view of these strong constraints, one might expect that no significant progress is possible at the LHC since DY cross sections, which are the best probes of Eq.~(\ref{eq:propagators}), are measured with at best a few percent accuracy \cite{Chatrchyan:2014mua,Aad:2016naf,CMS:2014jea,Aad:2016zzw}. This expectation is correct for ${\hat{\textrm{S}}}$ and ${\hat{\textrm{T}}}$, which only appear on the pole of the propagator, which is better constrained at LEP\@.  However, $\W$ and $\Y$ introduce constant terms in the propagator, modifying the cross sections by a factor that grows with energy as $q^2/\mw^2$.  Neutral DY measurements from the 8 TeV LHC~\cite{CMS:2014jea,Aad:2016zzw} have already achieved $10\%$ accuracy at a center of mass energy $q\sim1$~TeV, where this enhancement factor is above $100$. They could thus be already sensitive to values of $\W$ and $\Y$ as small as $10^{-3}$, outside the reach of LEP\@. Moreover, current high-energy measurements are statistics-dominated, the systematic component of the error being as small as $2\%$. Big improvements are thus possible at 13 TeV thanks to higher energy and luminosity.

The electroweak gauge boson propagators are modified by an effective Lagrangian, ${\mathcal{L}}$, containing the two dimension-6 operators from the middle column of Table~\ref{tab:op}\@.  These operators generate the $\W$ and $\Y$ parameters of Eq.~(\ref{eq:propagators}).  The effects of $\W$ and $\Y$ on DY are also captured by ${\mathcal{L}}'$, which consists of the operators from the right column of Table~\ref{tab:op}\@.    Here, $J_L$ and $J_Y$ are the SU$(2)_L$ and U$(1)_{Y}$ currents, and $g_{1,2}$  are the corresponding couplings.  The current bilinears contain quark-lepton contact operators (a subset of those considered in Ref.~\cite{deBlas:2013qqa})
which directly contribute to the DY amplitude with a term that grows with the energy, mimicking the effect of the modified propagators in Eq.~(\ref{eq:propagators}). 
The effective Lagrangian ${\mathcal{L}}'$ is obtained from ${\mathcal{L}}$ by field redefinitions, after truncating operators that are higher order in $\W$ and $\Y$ and with more derivatives.  ${\mathcal{L}}$ and 
${\mathcal{L}}'$ are physically inequivalent because of this truncation, however they agree in the limits of small $\W$ and $\Y$ and/or low energy.

\begin{figure*}[tb]
\centering
\includegraphics[width=0.88\linewidth]{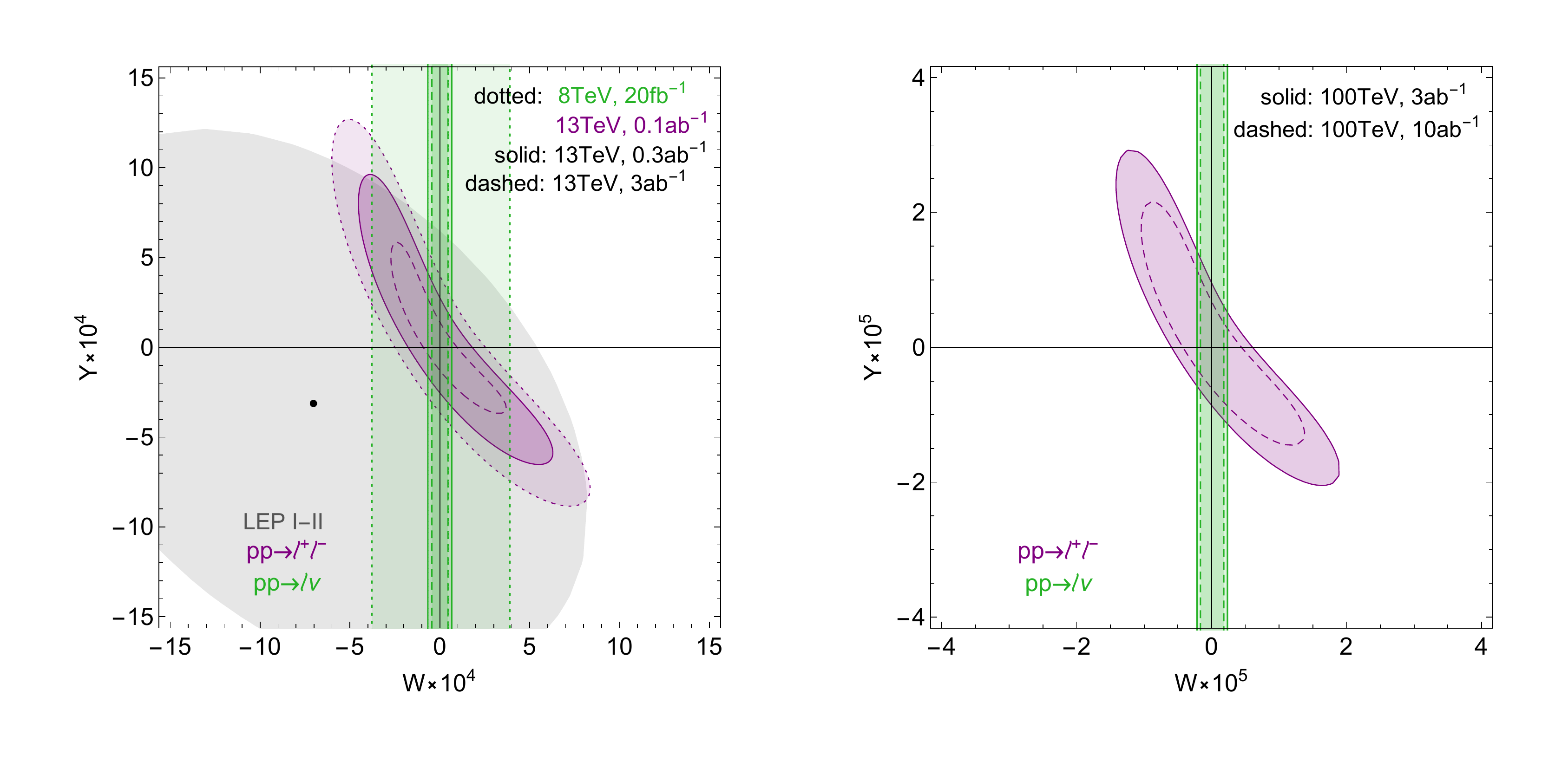}
\vspace{-.3cm}
\caption{
\label{fig1}
\it
Projected 95\% CL exclusions in the \W-\Y \,plane. \textbf{Left:}~ exclusion from neutral (purple) and charged (green) DY from LHC measurements at various luminosities and energies, compared to LEP bounds (gray). \textbf{Right:}~ projected reach from a 100\,TeV collider (notice the change of scale).}
\label{fig:prospects}
\end{figure*}

\noindent {\bf  Current Limits  and Future Prospects.---}
We compute the tree-level neutral ($pp\rightarrow l^+ l^-$) and charged ($pp\rightarrow l\nu$) DY differential cross sections with the modified propagators of Eq.~(\ref{eq:propagators}). The differential distribution is integrated in dilepton invariant mass (for neutral DY) and transverse lepton mass (for charged DY) bins and compared with the observations using a $\chi^2$ test. 
The value of the cross section in each bin can be written as $\sigma=\sigma_{SM}(1+\sum_p a_p C_p+\sum_{pq}b_{pq} C_pC_q)$, $C=\{\W,\Y\}$, and $a_p$, $b_{pq}$ are numbers that vary bin-by-bin. The coefficients $a_p$  represent the interference between the SM and the new physics, which is the leading effect in our case.
The SM cross section, $\sigma_{SM}$, is computed at NNLO QCD using FEWZ~\cite{Melnikov:2006kv,Catani:2009sm,Gavin:2010az,Li:2012wna,Catani:2007vq,Gavin:2012sy}. The NNPDF2.3@NNLO PDF~\cite{Ball:2012cx,Ball:2013hta}, with $\alpha_s=0.119$, is employed for the central value predictions at 8 and  13\,TeV, and to quantify PDF uncertainties. We use NNPDF3.0@NNLO~\cite{Ball:2014uwa} for 100\,TeV projections. The QCD scale and PDF uncertainties are included following Ref.~\cite{Alves:2014cda}. The photon PDF is not a significant source of uncertainty, because it was recently determined with high precision~\cite{Manohar:2016nzj}.

Run-$1$ limits on $\W$ and $\Y$ from neutral DY are obtained using the differential cross section measurements performed by ATLAS \cite{Aad:2016zzw} and CMS \cite{CMS:2014jea}, including the full correlation matrix of experimental uncertainties.
The left panel of Fig.~\ref{fig:data} shows the comparison of the ATLAS and CMS measurements with our theoretical predictions for the cross section in each bin in the SM ($\W=\Y=0$) hypothesis. Theoretical uncertainties from PDF and scale uncertainty are displayed as a shaded band, while the black error bars represent  experimental uncertainties. 
Our predictions reproduce observations, under the SM hypothesis, over the whole invariant mass range. We also notice that statistical errors are by far dominant at high mass, the theoretical and systematical uncertainties being one order of magnitude smaller, around $2\%$. The right panel of Fig.~\ref{fig:data} shows the $95\%$ exclusion contours obtained with ATLAS and CMS data in the $\W$-$\Y$ plane. The constraint from LEP and from other low-energy measurements~\cite{Falkowski:2015krw} is displayed as a grey region (marginalizing over $\hat{\textrm{S}}$ and $\hat{\textrm{T}}$). Run-$1$ limits from neutral DY are already competitive with LEP constraints.

\begin{figure*}[tb]
\centering
\includegraphics[width=1\linewidth]{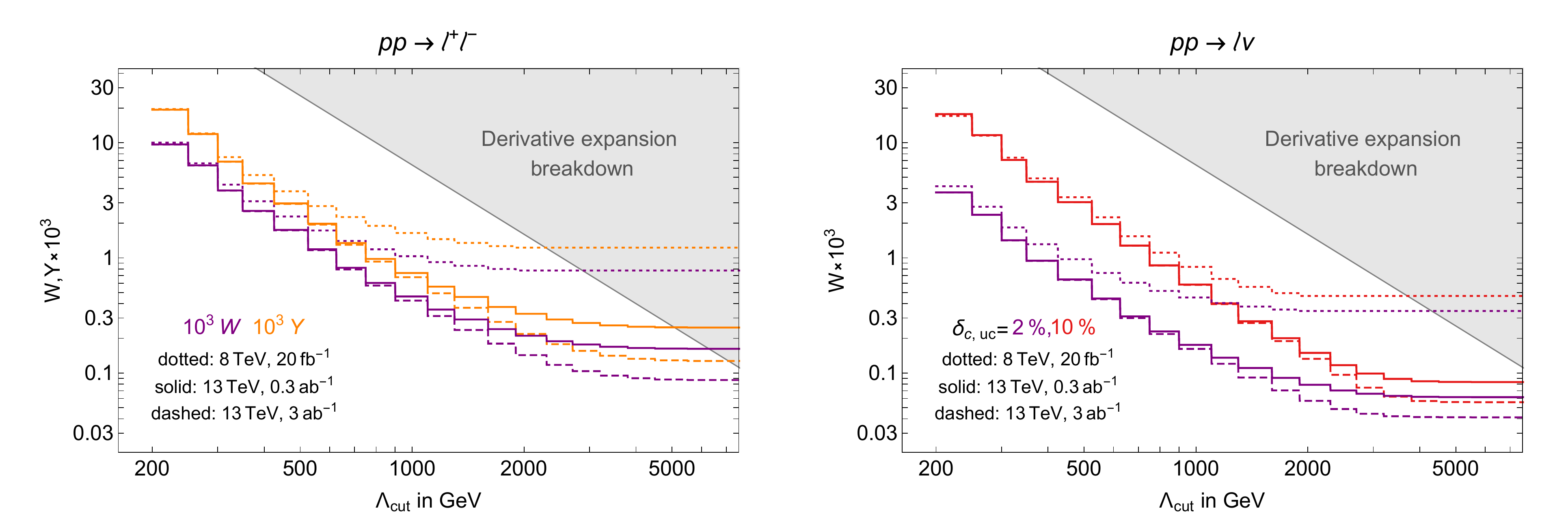}
\vspace{-.3cm}
\caption{
\label{fig1}
\em
Projected bounds as a function of a cutoff on the mass variable. The gray region corresponds to  $\Lambda_{\rm{cut}} > \Lambda_{\rm{max}}$ from Eq.~\ref{eq:maxcut}. \textbf{Left:}~ Bounds on \W (with $\Y=0$) or \Y (with $\W=0$) from neutral DY including only events with the dilepton invariant mass smaller than $\Lambda_{\rm{cut}}$. \textbf{Right:}~ Bounds on \W ~from charged DY including only events with the lepton transverse mass smaller than $\Lambda_{\rm{cut}}$. }
\label{fig:bins}
\end{figure*}

We project neutral/charged DY reach at $13$~TeV and at a future $100$~TeV collider.  We also project the reach of $8$~TeV for charged DY (differential cross section measurements are presently unavailable at high transverse mass). In order to estimate experimental uncertainties, we include fully correlated ($\delta_{c}$) and uncorrelated ($\delta_{uc}$) uncertainties. For neutral DY, we use $\delta_c=\delta_{uc}=2\%$, commensurate with uncertainties achieved in existing 8 TeV measurements.  For charged DY we use $\delta_c=\delta_{uc}=5\%$, consistent with uncertainty attributed to charged DY backgrounds to $W'$ searches~\cite{ATLAS:2014wra, Khachatryan:2014tva,Aaboud:2016zkn}.  We apply the cuts $p^{\ell}_T>25$~GeV and $|\eta_\ell|<2.5$ on leptons, and assume an identification efficiency of $65\%$ ($80\%$) for electrons (muons).   For neutral (charged) DY we bin invariant (transverse) mass as in Ref.~\cite{Alves:2014cda}. 

Our $13$~TeV results, overlaid with the LEP limit, are shown in Fig.~\ref{fig:prospects} left, for luminosities of $100, 300$, and $3000$~fb$^{-1}$.
The projected LHC limits are radically better than present constraints.
The expected Run-$1$ limit on $\W$ from charged DY is shown as a dotted green band. The reach far surpasses LEP, even with Run-$1$ data. Projections for $100$\,TeV are shown to the right of Fig.~\ref{fig:prospects} for luminosities of $3$ and $10$~ab$^{-1}$.

\begin{figure*}[tb]
\centering
\includegraphics[width=1\linewidth]{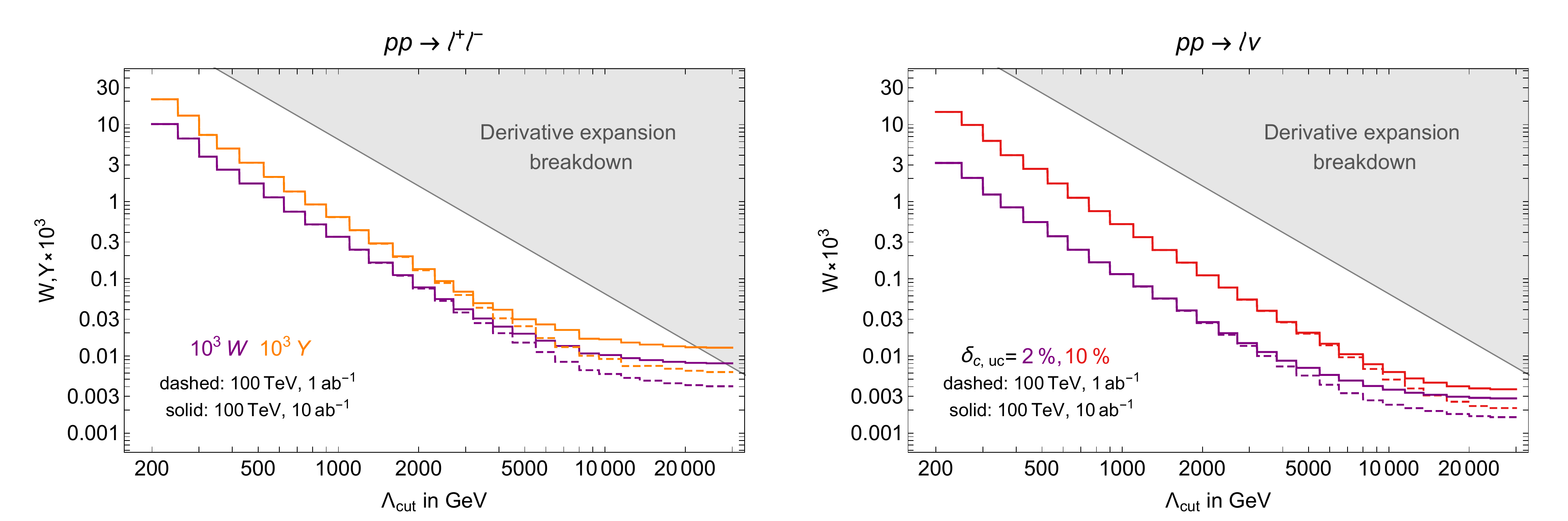}
\vspace{-.3cm}
\caption{
\label{fig1}
\em
Projected bounds as a function of a cutoff on the mass variable for a $pp$ collider with 100\,TeV center of mass energy. The bounds are plotted as in Fig.~\ref{fig:bins}.}
\label{fig:bins_100}
\end{figure*}

In order to delve deeper into our results, Fig.~\ref{fig:bins} shows how the limit on $\W$ or $\Y$ changes if only invariant mass (for neutral DY, left panel) or transverse mass (for charged DY, right panel) bins below a certain threshold $\Lambda_{\rm{cut}}$ are included.  We learn that our limits mainly rely on measurements below $1~(2)$~TeV for $\sqrt s = 8~(13)$~TeV\@.  The dramatic improvement of reach with $\sqrt s$ is a direct consequence of how the relevant bins scale with $\sqrt s$, as visible in Fig.~\ref{fig:bins}, leading to an improvement of sensitivity to $\W$ or $\Y$ that scales as $q^2/\mw^2 \propto s$.  By highlighting the relevant bins,  Fig.~\ref{fig:bins} illustrates 
the ranges of invariant/transverse mass where percent-level experimental systematics will be important.  The effect of varying the systematic uncertainties down ($2\%$) or up ($10\%$) with respect to our estimate (i.e., $5\%$ for charged DY) is shown on the right panel of Fig.~\ref{fig:bins}. Similar bounds but for a 100\,TeV center of mass $pp$ collider are shown in Fig~\ref{fig:bins_100}. In this case the plots show that the bounds mainly rely on invariant mass measurements (transverse mass measurements in the case of charged DY) below 10\,TeV.

\begin{table*}[t]
\begin{center}
{\small
\begin{tabular}{c|c|c|c|c|c|c|c|c|c|c|c}  
 \multicolumn{2}{c|}{  }   & LEP& ATLAS\,8 & CMS\,8 & \multicolumn{2}{c |}{ LHC\,13}  & 100\,TeV &  ILC  & TLEP &~ CEPC ~  &~ ILC 500\,GeV~  \\ \hline \hline
 
 \multicolumn{2}{ r|}{ luminosity \hspace{4cm}}  & ~~$2\times10^7\,Z$~~& ~$19.7\,\text{fb}^{-1}$~ & ~$20.3\,\text{fb}^{-1}$~ & ~$0.3\,\text{ab}^{-1}$ ~& ~ $3\,\text{ab}^{-1}$~&~ $10\,\text{ab}^{-1}$~&  ~$10^9\,Z$ ~&  ~$10^{12}\,Z$~  &  ~$10^{10}\,Z$~ &~ $3\,\text{ab}^{-1}$~\\ \hline\hline
 
 ~NC~  & ~\W $\times 10^{4}$~  & $[-19, 3]$ & $[-3,15]$  & $[-5,22]$  & $\pm1.5$ & $\pm0.8$  & $\pm0.04$  &  $\pm 4.2$  & $\pm 1.2$ &$\pm3.6$ &$\pm0.3$\\ 
 \cline{2-12}
  & ~\Y $\times 10^{4}$~ & $[-17, 4]$  & $[-4,24]$  & $[-7,41]$  & $\pm2.3$ & $\pm1.2$  & $\pm0.06$  & $\pm1.8$  & $\pm1.5$&$\pm3.1$  &$\pm0.2$ \\ \hline
~CC~  & \W  $\times 10^{4}$~& ---  &\multicolumn{2}{ c |}{ $\pm 3.9$}  & $\pm0.7$ & $\pm0.45$  & $\pm0.02$   & ---  &   --- & ---  & ---  \\ 
\end{tabular}
}
\end{center}
  \caption{ \label{tab:boost} \it
  Reach on \W and \Y~from different machines with various energies and luminosities (95\% CL). The bounds from neutral DY are obtained setting the unconstrained parameter to zero. Bounds from LEP are extracted from \cite{Falkowski:2015krw}, marginalizing over $\hat{\rm{S}}$ and $\hat{{\rm T}}$. Bounds from $Z$-peak {\textrm{ILC}}~\cite{Baer:2013cma}, {\textrm{TLEP}}~\cite{Gomez-Ceballos:2013zzn} and {\textrm{CEPC}}~\cite{CEPC-SPPCStudyGroup:2015csa} are from Ref.~\cite{Fan:2014axa}. Bounds from off-peak measurements of $e^+e^-\to e^+e^-$ at lepton colliders are extracted from \cite{Harigaya:2015yaa}.
  }
\end{table*}

The shape of the limit/reach contours in the $\W$-$\Y$ plane can be understood as follows.    The interference term in the partonic neutral DY cross section depends on a $q^2$-independent linear combination of $\W$ and $\Y$, when integrated over angles.   The orthogonal combination is only constrained when $\W$ and $\Y$ are large enough for quadratic terms to be relevant.  In view of the strong constraint expected on $\W$ from charged DY, this flat direction is irrelevant in practice.  However, we note that the flat direction can in principle be constrained with neutral DY only, using angular information such as the energy dependence of  forward-backward asymmetries~\cite{deBlas:2013qqa}.  In practice, this does not improve the 8\,TeV limits (due to the dominance of the $q_L{\overline{q}}_R\rightarrow l^-_Ll^+_R$ amplitude), but may be significant at higher energies/luminosities.  We leave a full study of the power of angular distributions to future work.

\noindent {\bf Beyond EFT's.---} When using EFTs to describe high energy processes, one has to keep in mind that an EFT provides an accurate description of the underlying new physics only at energies below the new physics scale. The latter scale is the EFT cutoff and it should be regarded as a free parameter of the EFT \cite{Racco:2015dxa}. A related concept is that of ``maximal cutoff'', which is the maximal new physics scale that can produce an EFT operator of a given magnitude (e.g., a given value of $\W$ or $\Y$). The EFT limits become inconsistent if they come from energies above the cutoff.
This concept has been addressed in DM EFT searches~\cite{Racco:2015dxa,Busoni:2013lha} and electroweak EFT studies~\cite{Contino:2016jqw}. Depending on whether we consider new physics that directly generates contact interactions ($\mathcal L'$), or modifies the vacuum polarizations ($\mathcal L$), the maximal cutoff estimate is,
\be\label{eq:maxcut}
\Lambda'\equiv\frac{4\pi m_W/g_2}{\max(\sqrt\W,t\sqrt\Y)} \,,~~\Lambda\equiv \frac{m_W}{\max(\sqrt\W,\sqrt\Y)}<\Lambda'\,.
\ee
The first estimate comes from demanding $2\rightarrow2$ amplitudes induced by ${\mathcal{L}}'$ not to exceed the $16\pi^2$ perturbativity bound, the second one from the validity of the derivative expansion, taking into account that ${\mathcal{L}}$ is a higher-derivative correction to the (canonically normalized) vector boson kinetic terms. There is no contradiction in the fact that the two pictures have different cutoffs  since ${\mathcal{L}}$ and ${\mathcal{L}}'$ are equivalent only if the $d>6$ operators induced by the field redefinition are negligible (as is the case when $q<\Lambda$). 

In order to quantify the impact of the limited EFT validity, Figs.~\ref{fig:bins} and \ref{fig:bins_100} shows how the reach deteriorates when only data below the cutoff are employed.\footnote{This is not completely correct in the charged DY case because low transverse mass bins might in principle still receive contributions from reactions that occur at very high center of mass energies, well above the cutoff. These contribution are however negligible for the analysis discussed in this paper. To show this we recalculated the bounds on $\W$ shown in Table~\ref{tab:boost} artificially including in the calculation of the New Physics cross section only those events in which the lepton-neutrino invariant mass is below the maximal cutoff $\Lambda=m_W/\sqrt{\W}$ at which the derivative expansion breaks down. These new bounds are only weaker than the old ones by a few percent, showing that the contamination mentioned above is numerically irrelevant.} If the resulting curve stays below the maximal cutoff lines corresponding to Eq.~(\ref{eq:maxcut}), as in our case, the EFT limit is self-consistent. The right panels of Figs.~\ref{fig:bins} and \ref{fig:bins_100} also show how lowering the systematic uncertainties moves the limit curve far from the maximal cutoff line. This allows to test EFTs with below maximal cutoffs.

Our results can be applied to various new physics scenarios.
Higher derivative corrections to the SM gauge boson kinetic terms directly test their compositeness above a scale $\Lambda_2\approx m_W/\sqrt{\W}$ for the SU(2) gauge fields and $\Lambda_1\approx m_W/\sqrt{\Y}$ for the hypercharge. Our results imply $\Lambda_2\gtrsim 4$\,TeV from charged DY at 8\,TeV and $(\Lambda_2,\Lambda_1)\gtrsim (6.5, 5)$\,TeV  from neutral DY with an LHC luminosity of 300~fb$^{-1}$\@. 
Our bounds are also applicable to models in which elementary $W^\pm$ and $B$ bosons mix with heavy vector resonances. To discuss the bound in a quantitative way we consider an SU(2)$_L$ triplet massive vector field, $V$, coupled to the $SU(2)_L$ current of the SM.  This matter content is described by the following effective Lagrangian:
\be\label{vectormodel}
\mathcal L_V=-\frac{1}{4} D_{[\mu}V^a_{\nu]}D^{[\mu}V^{a\nu]}+\frac{M^2}{2}V^a_\mu V^{a\mu}-g_V V^{a\mu} J_\mu^a,
\ee
where we define the covariant derivative for $V$ as $D_\mu V^a_\nu\equiv \partial_\mu V^a_\nu+g \epsilon^{abc} W_\mu^b V_\nu^c$ and the $SU(2)_L$ current $J^a_\mu$ as
\be
J^{a\mu}=\sum_f \bar f_L\gamma^\mu \tau^a f_L+iH^\dagger \tau^a \overleftrightarrow{D}^\mu H
\ee
with $f$ running over SM quarks and leptons and $H$ being the Higgs boson doublet.
Ref.~\cite{Pappadopulo:2014qza} discusses possible UV realizations of Eq.~(\ref{vectormodel}): the vector field $V$ can either belong to a weakly coupled UV completion or it can be a composite resonance as those arising in models of Higgs compositeness. Integrating out the vector triplet generates $\W$ as follows,
\be\label{Wvector}
\W=\frac{g_V^2}{g^2}\frac{m_W^2}{M^2}+O(\W^2).
\ee

\begin{figure*}[tb]
\centering
\includegraphics[width=0.88\linewidth]{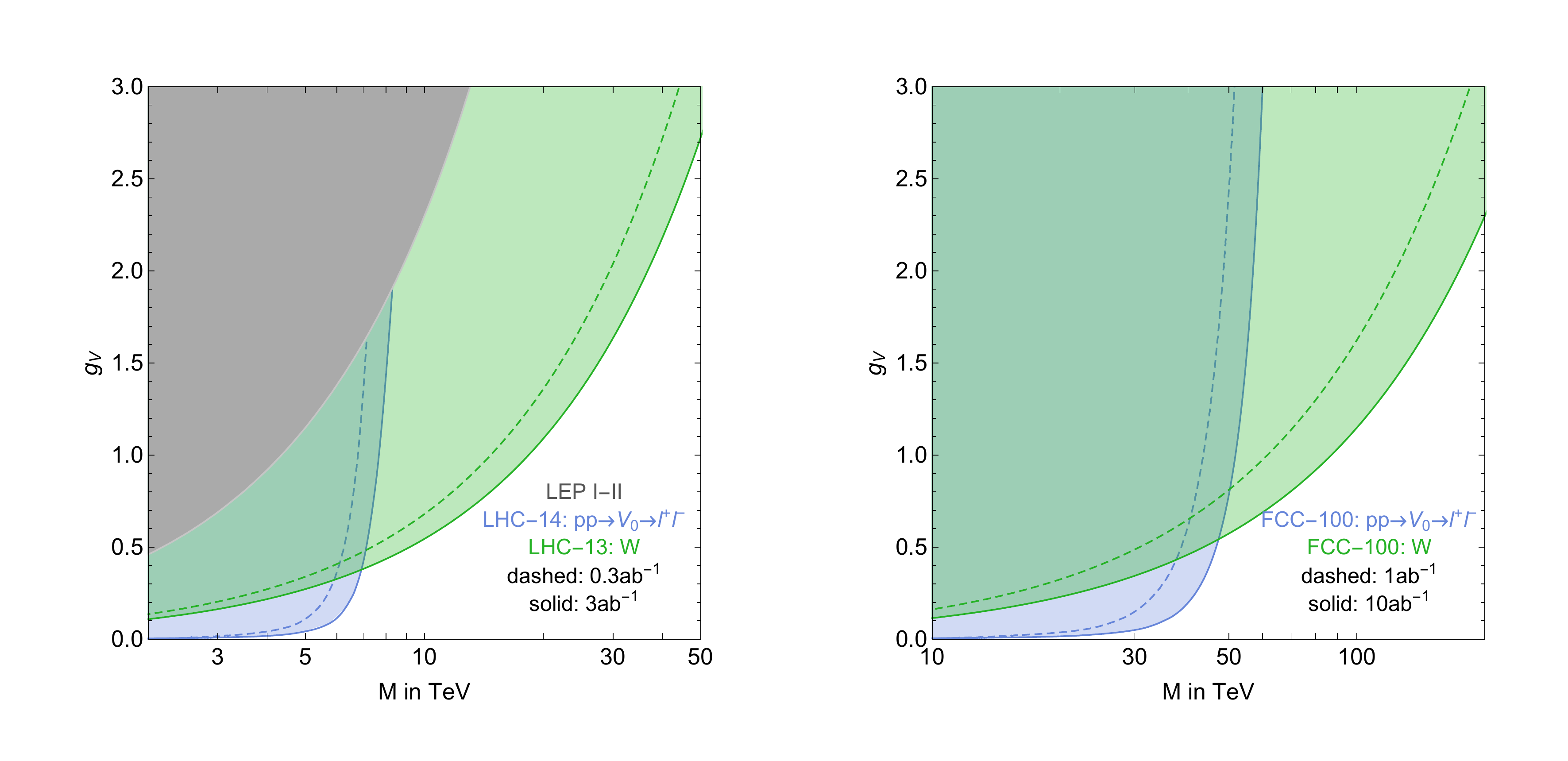}
\vspace{-.3cm}
\caption{
\label{fig1}
\em
Projected experimental reach for the vector model of Eq.~(\ref{vectormodel}). \textbf{Left:}~ Projected bounds from the LHC coming from our DY constraint on $\W$ (green, 13 TeV) and direct searches $pp\to V_3\to\ell^+\ell^-$ (blue, 14 TeV).  Solid (dashed) lines correspond to 3 (0.3) ab$^{-1}$.  For comparison, the LEP bound on $\W$ is shown in gray.  \textbf{Right:} Projected bounds from a 100 TeV FCC from our DY constraint on $\W$ (green) and direct searches (blue).}
\label{fig:vectormodel}
\end{figure*}

The model in Eq.~(\ref{vectormodel}) is described by two parameters $M$ and $g_V$. In Fig.~\ref{fig:vectormodel} we show the bounds on the model in the $(M,g_V)$ plane coming from $\W$, see Eq.~(\ref{Wvector}). We use the current and projected sensitivity of LHC and a 100\,TeV Future Circular Collider (FCC) to $pp\to V_3\to\ell^+\ell^-$ as extracted from Ref.~\cite{Thamm:2015zwa}.


\noindent {\bf  Outlook.---} 
In this paper, we have demonstrated that hadron colliders can be used to perform electroweak precision tests, and in particular that the LHC is now surpassing LEP in sensitivity to the universal parameters $\W$ and $\Y$.  Our results are summarized in Table~\ref{tab:boost}, where we also compare to future lepton colliders. 

We conclude by noting that the universal parameters $\W$ and $\Y$ are just two examples from the class of operators of the SM EFT whose effects grow with energy.  The LHC, and future hadron colliders, therefore have great potential to perform {\it precision} tests, because high center of mass energy compensates limited accuracy.  We advocate exploration of a  broad program of precision tests at hadron colliders, where SM measurements can be leveraged as indirect probes of new physics that is too heavy to produce directly.

\newpage
\vspace{.3cm}
\begin{acknowledgements}
\noindent {\bf \em Acknowledgements.---}We thank Daniele Alves, Kyle Cranmer, Jorge de Blas, Roberto Contino, Jamison Galloway, Andy Haas, Alex Pomarol, Francesco Riva, and Jon Walsh for helpful discussions.  G.~P., R.~T. and A.~W.~thank Javi Serra for collaboration at an early stage of this project. We thank the authors of Ref.~\cite{Fan:2014axa} for providing the code we used to compute the bounds on $\W$ and  $\Y$ at future lepton colliders.
M.~F. is supported by the DOE grant DOE-SC0010008. The work of G.~P. is partly supported by MINECO under Grant CICYT- FEDER-FPA2014-55613-P, by the Severo Ochoa Excellence Program of MINECO under the grant SO-2012-0234 and by Secretaria d'€™Universitats i Recerca del Departament dâ'™Economia i Coneixement de la Generalitat de Catalunya under Grant 2014 SGR 1450. D.~P. is supported by the James Arthur Postdoctoral Fellowship. J.~T.~R. is supported by the NSF CAREER grant PHY-1554858, and acknowledges  the hospitality of the Aspen Center for Physics, supported by NSF grant PHY-1066293. The work of R.~T. is supported by the Swiss National Science Foundation under grants CRSII2-160814 and 200020-150060. A.~W. acknowledges the MIUR-FIRB grant RBFR12H1MW and the ERC Advanced Grant no. 267985 (DaMeSyFla).
\end{acknowledgements}

\bibliographystyle{mine}
\bibliography{DY}

\providecommand{\href}[2]{#2}\begingroup\raggedright\begin{thebibliography}{10}

\bibitem{ALEPH:2005ab}
{\bf ALEPH}, {\bf DELPHI}, {\bf L3}, {\bf OPAL}, {\bf SLD}, {\bf LEP
  Electroweak Working Group}, {\bf SLD Electroweak Group}, {{\bf SLD Heavy
  Flavour Group} Collaborations}, S.~Schael {\em et al.,}
  \href{http://dx.doi.org/10.1016/j.physrep.2005.12.006}{{\em Phys. Rept.}
  {\bfseries 427} (2006) 257--454},
  \href{http://arxiv.org/abs/hep-ex/0509008}{{\tt hep-ex/0509008}}.
  [\href{http://inspirehep.net/record/691576}{Inspire}].

\bibitem{Domenech:2012ai}
O.~Domenech, A.~Pomarol and J.~Serra,
  \href{http://dx.doi.org/10.1103/PhysRevD.85.074030}{{\em Phys. Rev.}
  {\bfseries D 85} (2012) 074030}, \href{http://arxiv.org/abs/1201.6510}{{\tt
  arXiv:1201.6510}}. [\href{http://inspirehep.net/record/1086831}{Inspire}].

\bibitem{Biekoetter:2014jwa}
A.~Biekoetter, A.~Knochel, M.~Kraemer, D.~Liu and F.~Riva,
  \href{http://dx.doi.org/10.1103/PhysRevD.91.055029}{{\em Phys. Rev.}
  {\bfseries D 91} (2015) 055029}, \href{http://arxiv.org/abs/1406.7320}{{\tt
  arXiv:1406.7320}}. [\href{http://inspirehep.net/record/1303909}{Inspire}].

\bibitem{Azatov:2015oxa}
A.~Azatov, R.~Contino, G.~Panico and M.~Son,
  \href{http://dx.doi.org/10.1103/PhysRevD.92.035001}{{\em Phys. Rev.}
  {\bfseries D 92} no.~3, (2015) 035001},
  \href{http://arxiv.org/abs/1502.00539}{{\tt arXiv:1502.00539}}.
  [\href{http://inspirehep.net/record/1342472}{Inspire}].

\bibitem{Dror:2015nkp}
J.~A. Dror, M.~Farina, E.~Salvioni and J.~Serra,
  \href{http://dx.doi.org/10.1007/JHEP01(2016)071}{{\em JHEP} {\bfseries 1601}
  (2016) 071}, \href{http://arxiv.org/abs/1511.03674}{{\tt arXiv:1511.03674}}.
  [\href{http://inspirehep.net/record/1404154}{Inspire}].

\bibitem{Butter:2016cvz}
A.~Butter, O.~J.~P. \'Eboli, J.~Gonzalez-Fraile, M.~C. Gonzalez-Garcia,
  T.~Plehn and M.~Rauc, \href{http://dx.doi.org/10.1007/JHEP07(2016)152}{{\em
  JHEP} {\bfseries 07} (2016) 152}, \href{http://arxiv.org/abs/1604.03105}{{\tt
  arXiv:1604.03105}}. [\href{http://inspirehep.net/record/1445112}{Inspire}].

\bibitem{Falkowski:2016cxu}
A.~Falkowski, M.~Gonzalez-Alonso, A.~Greljo, D.~Marzocca and M.~Son,
  \href{http://arxiv.org/abs/1609.06312}{{\tt arXiv:1609.06312}}.
  [\href{http://inspirehep.net/record/1487539}{Inspire}].

\bibitem{Buchmuller:1985jz}
W.~Buchmuller and D.~Wyler,
\href{http://dx.doi.org/10.1016/0550-3213(86)90262-2}{{\em Nucl. Phys.}
  {\bfseries B268} (1986) 621--653}.

\bibitem{Grzadkowski:2010es}
B.~Grzadkowski, M.~Iskrzynski, M.~Misiak and J.~Rosiek,
  \href{http://dx.doi.org/10.1007/JHEP10(2010)085}{{\em JHEP} {\bfseries 10}
  (2010) 085},
\href{http://arxiv.org/abs/1008.4884}{{\tt arXiv:1008.4884}}.

\bibitem{Barbieri:2004qk}
R.~Barbieri, A.~Pomarol, R.~Rattazzi and A.~Strumia,
  \href{http://dx.doi.org/10.1016/j.nuclphysb.2004.10.014}{{\em Nucl. Phys.}
  {\bfseries B 703} (2004) 127},
  \href{http://arxiv.org/abs/hep-ph/0405040}{{\tt hep-ph/0405040}}.
  [\href{http://inspirehep.net/record/649700}{Inspire}].

\bibitem{Wells:2015uba}
J.~D. Wells and Z.~Zhang, \href{http://dx.doi.org/10.1007/JHEP01(2016)123}{{\em
  JHEP} {\bfseries 01} (2016) 123},
\href{http://arxiv.org/abs/1510.08462}{{\tt arXiv:1510.08462}}.

\bibitem{Wells:2015cre}
J.~D. Wells and Z.~Zhang, \href{http://dx.doi.org/10.1007/JHEP06(2016)122}{{\em
  JHEP} {\bfseries 06} (2016) 122},
\href{http://arxiv.org/abs/1512.03056}{{\tt arXiv:1512.03056}}.

\bibitem{Drell:1970wh}
S.~D. Drell and T.-M. Yan,
  \href{http://dx.doi.org/10.1103/PhysRevLett.25.316}{{\em Phys. Rev. Lett.}
  {\bfseries 25} (1970) 316--320}. [Erratum: Phys. Rev. Lett.25,902(1970)].
  [\href{http://inspirehep.net/record/60911}{Inspire}].

\bibitem{Salvioni:2009mt}
E.~Salvioni, G.~Villadoro and F.~Zwirner,
  \href{http://dx.doi.org/10.1088/1126-6708/2009/11/068}{{\em JHEP} {\bfseries
  0911} (2009) 068}, \href{http://arxiv.org/abs/0909.1320}{{\tt
  arXiv:0909.1320}}. [\href{http://inspirehep.net/record/830532}{Inspire}].

\bibitem{Contino:2010mh}
R.~Contino, C.~Grojean, M.~Moretti, F.~Piccinini and R.~Rattazzi,
  \href{http://dx.doi.org/10.1007/JHEP05(2010)089}{{\em JHEP} {\bfseries 05}
  (2010) 089}, \href{http://arxiv.org/abs/1002.1011}{{\tt arXiv:1002.1011}}.
  [\href{http://inspirehep.net/record/845195}{Inspire}].

\bibitem{delAguila:2010mx}
F.~del Aguila, J.~de~Blas and M.~Perez-Victoria,
  \href{http://dx.doi.org/10.1007/JHEP09(2010)033}{{\em JHEP} {\bfseries 09}
  (2010) 033}, \href{http://arxiv.org/abs/1005.3998}{{\tt arXiv:1005.3998}}.
  [\href{http://inspirehep.net/record/855936}{Inspire}].

\bibitem{deBlas:2012qp}
J.~de~Blas, J.~M. Lizana and M.~Perez-Victoria,
  \href{http://dx.doi.org/10.1007/JHEP01(2013)166}{{\em JHEP} {\bfseries 01}
  (2013) 166}, \href{http://arxiv.org/abs/1211.2229}{{\tt arXiv:1211.2229}}.
  [\href{http://inspirehep.net/record/1201947}{Inspire}].

\bibitem{Lizana:2013xla}
J.~M. Lizana and M.~Perez-Victoria,
  \href{http://dx.doi.org/10.1051/epjconf/20136017008}{{\em Eur. Phys. J. Web
  Conf.} {\bfseries 60} (2013) 17008},
  \href{http://arxiv.org/abs/1307.2589}{{\tt arXiv:1307.2589}}.
  [\href{http://inspirehep.net/record/1242121}{Inspire}].

\bibitem{Pappadopulo:2014qza}
D.~Pappadopulo, A.~Thamm, A.~Thamm, A.~Wulzer and R.~Torre,
  \href{http://dx.doi.org/10.1007/JHEP09(2014)060}{{\em JHEP} {\bfseries 1409}
  (2014) 060}, \href{http://arxiv.org/abs/1402.4431}{{\tt arXiv:1402.4431}}.
  [\href{http://inspirehep.net/record/745577}{Inspire}].

\bibitem{Cirelli:2005uq}
M.~Cirelli, N.~Fornengo and A.~Strumia,
  \href{http://dx.doi.org/10.1016/j.nuclphysb.2006.07.012}{{\em Nucl. Phys.}
  {\bfseries B 753} (2006) 178},
  \href{http://arxiv.org/abs/hep-ph/0512090}{{\tt hep-ph/0512090}}.
  [\href{http://inspirehep.net/record/699850}{Inspire}].

\bibitem{Liu:2016idz}
D.~Liu, A.~Pomarol, R.~Rattazzi and F.~Riva,
  \href{http://arxiv.org/abs/1603.03064}{{\tt arXiv:1603.03064}}.
  [\href{http://inspirehep.net/record/1427033}{Inspire}].

\bibitem{Peskin:1990zt}
M.~E. Peskin and T.~Takeuchi,
  \href{http://dx.doi.org/10.1103/PhysRevLett.65.964}{{\em Phys. Rev. Lett.}
  {\bfseries 65} (1990) 964}.
  [\href{http://inspirehep.net/record/296528}{Inspire}].

\bibitem{Chatrchyan:2011nv}
{\bfseries CMS} Collaboration, S.~Chatrchyan {\em et al.,}
  \href{http://dx.doi.org/10.1007/JHEP08(2011)117}{{\em JHEP} {\bfseries 1108}
  (2011) 117}, \href{http://arxiv.org/abs/1104.1617}{{\tt arXiv:1104.1617}}.
  [\href{http://inspirehep.net/record/895581}{Inspire}].

\bibitem{CMS:2011aa}
{\bfseries CMS} Collaboration, S.~Chatrchyan {\em et al.,}
  \href{http://dx.doi.org/10.1007/JHEP10(2011)132}{{\em JHEP} no.~1110, (2011)
  132}, \href{http://arxiv.org/abs/1107.4789}{{\tt arXiv:1107.4789}}.
  [\href{http://inspirehep.net/record/919737}{Inspire}].

\bibitem{Aad:2011dm}
{\bfseries ATLAS} Collaboration, G.~Aad {\em et al.,}
  \href{http://dx.doi.org/10.1103/PhysRevD.85.072004}{{\em Phys. Rev.}
  {\bfseries D 85} (2012) 072004}, \href{http://arxiv.org/abs/1109.5141}{{\tt
  arXiv:1109.5141}}. [\href{http://inspirehep.net/record/928289}{Inspire}].

\bibitem{Chatrchyan:2014mua}
{\bfseries CMS} Collaboration, S.~Chatrchyan {\em et al.,}
  \href{http://dx.doi.org/10.1103/PhysRevLett.112.191802}{{\em Phys. Rev.
  Lett.} {\bfseries 112} (2014) 191802},
  \href{http://arxiv.org/abs/1402.0923}{{\tt arXiv:1402.0923}}.
  [\href{http://inspirehep.net/record/1280200}{Inspire}].

\bibitem{CMS:2014jea}
{\bfseries CMS} Collaboration, V.~Khachatryan {\em et al.,}
  \href{http://dx.doi.org/10.1140/epjc/s10052-015-3364-2}{{\em Eur. Phys. J.}
  {\bfseries C 75} no.~4, (2015) 147},
  \href{http://arxiv.org/abs/1412.1115}{{\tt arXiv:1412.1115}}.
  [\href{http://inspirehep.net/record/1332509}{Inspire}].

\bibitem{Aad:2016naf}
{\bfseries ATLAS} Collaboration, G.~Aad {\em et al.,}
  \href{http://dx.doi.org/10.1016/j.physletb.2016.06.023}{{\em Phys. Lett.}
  {\bfseries B 759} (2016) 601}, \href{http://arxiv.org/abs/1603.09222}{{\tt
  arXiv:1603.09222}}. [\href{http://inspirehep.net/record/1436497}{Inspire}].

\bibitem{Aad:2016zzw}
{\bfseries ATLAS} Collaboration, G.~Aad {\em et al.,}
  \href{http://dx.doi.org/10.1007/JHEP08(2016)009}{{\em JHEP} {\bfseries 1608}
  (2016) 009}, \href{http://arxiv.org/abs/1606.01736}{{\tt arXiv:1606.01736}}.
  [\href{http://inspirehep.net/record/1467454}{Inspire}].

\bibitem{Hamberg:1990np}
R.~Hamberg, W.~van Neerven and T.~Matsuura,
  \href{http://dx.doi.org/10.1016/S0550-3213(02)00814-3}{{\em Nucl. Phys.}
  {\bfseries B 359} (1991) 343, Erratum:
  \href{http://dx.doi.org/10.1016/0550--3213(91)90064--5}{Nucl. Phys. B {\bf
  644} (2002) 403}}. [\href{http://inspirehep.net/record/300492}{Inspire}].

\bibitem{Anastasiou:2003yy}
C.~Anastasiou, L.~J. Dixon, K.~Melnikov and F.~Petriello,
  \href{http://dx.doi.org/10.1103/PhysRevLett.91.182002}{{\em Phys. Rev. Lett.}
  {\bfseries 91} (2003) 182002},
  \href{http://arxiv.org/abs/hep-ph/0306192}{{\tt hep-ph/0306192}}.
  [\href{http://inspirehep.net/record/621730}{Inspire}].

\bibitem{Anastasiou:2003ds}
C.~Anastasiou, L.~J. Dixon, K.~Melnikov and F.~Petriello,
  \href{http://dx.doi.org/10.1103/PhysRevD.69.094008}{{\em Phys. Rev.}
  {\bfseries D 69} (2004) 094008},
  \href{http://arxiv.org/abs/hep-ph/0312266}{{\tt hep-ph/0312266}}.
  [\href{http://inspirehep.net/record/635943}{Inspire}].

\bibitem{Melnikov:2006kv}
K.~Melnikov and F.~Petriello,
  \href{http://dx.doi.org/10.1103/PhysRevD.74.114017}{{\em Phys. Rev.}
  {\bfseries D 74} (2006) 114017},
  \href{http://arxiv.org/abs/hep-ph/0609070}{{\tt hep-ph/0609070}}.
  [\href{http://inspirehep.net/record/725573}{Inspire}].

\bibitem{Catani:2009sm}
S.~Catani, L.~Cieri, G.~Ferrera, D.~de~Florian and M.~Grazzini,
  \href{http://dx.doi.org/10.1103/PhysRevLett.103.082001}{{\em Phys. Rev.
  Lett.} {\bfseries 103} (2009) 082001},
  \href{http://arxiv.org/abs/0903.2120}{{\tt arXiv:0903.2120}}.
  [\href{http://inspirehep.net/record/815316}{Inspire}].

\bibitem{Gavin:2010az}
R.~Gavin, Y.~Li, F.~Petriello and S.~Quackenbush,
  \href{http://dx.doi.org/10.1016/j.cpc.2011.06.008}{{\em Comput. Phys.
  Commun.} {\bfseries 182} (2011) 2388},
  \href{http://arxiv.org/abs/1011.3540}{{\tt arXiv:1011.3540}}.
  [\href{http://inspirehep.net/record/877524}{Inspire}].

\bibitem{Li:2012wna}
Y.~Li and F.~Petriello,
  \href{http://dx.doi.org/10.1103/PhysRevD.86.094034}{{\em Phys. Rev.}
  {\bfseries D 86} (2012) 094034}, \href{http://arxiv.org/abs/1208.5967}{{\tt
  arXiv:1208.5967}}. [\href{http://inspirehep.net/record/1182519}{Inspire}].

\bibitem{Cirigliano:2012ab}
V.~Cirigliano, M.~Gonzalez-Alonso and M.~L. Graesser,
  \href{http://dx.doi.org/10.1007/JHEP02(2013)046}{{\em JHEP} {\bfseries 02}
  (2013) 046},
\href{http://arxiv.org/abs/1210.4553}{{\tt arXiv:1210.4553}}.

\bibitem{Aad:2012bsa}
{\bfseries ATLAS} Collaboration, G.~Aad {\em et al.,}
  \href{http://dx.doi.org/10.1103/PhysRevD.87.015010}{{\em Phys. Rev.}
  {\bfseries D 87} no.~1, (2013) 015010},
  \href{http://arxiv.org/abs/1211.1150}{{\tt arXiv:1211.1150}}.
  [\href{http://inspirehep.net/record/1198341}{Inspire}].

\bibitem{Chatrchyan:2012hda}
{\bfseries CMS} Collaboration, S.~Chatrchyan {\em et al.,}
  \href{http://dx.doi.org/10.1103/PhysRevD.87.032001}{{\em Phys. Rev.}
  {\bfseries D 87} no.~3, (2013) 032001},
  \href{http://arxiv.org/abs/1212.4563}{{\tt arXiv:1212.4563}}.
  [\href{http://inspirehep.net/record/1208097}{Inspire}].

\bibitem{Chatrchyan:2013lga}
{\bfseries CMS} Collaboration, S.~Chatrchyan {\em et al.,}
  \href{http://dx.doi.org/10.1103/PhysRevD.87.072005}{{\em Phys. Rev.}
  {\bfseries D87} no.~7, (2013) 072005},
\href{http://arxiv.org/abs/1302.2812}{{\tt arXiv:1302.2812}}.

\bibitem{deBlas:2013qqa}
J.~de~Blas, M.~Chala and J.~Santiago,
  \href{http://dx.doi.org/10.1103/PhysRevD.88.095011}{{\em Phys. Rev.}
  {\bfseries D 88} (2013) 095011}, \href{http://arxiv.org/abs/1307.5068}{{\tt
  arXiv:1307.5068}}. [\href{http://inspirehep.net/record/1243596}{Inspire}].

\bibitem{Aad:2014wca}
{\bfseries ATLAS} Collaboration, G.~Aad {\em et al.,}
  \href{http://dx.doi.org/10.1140/epjc/s10052-014-3134-6}{{\em Eur. Phys. J.}
  {\bfseries C 74} no.~12, (2014) 3134},
  \href{http://arxiv.org/abs/1407.2410}{{\tt arXiv:1407.2410}}.
  [\href{http://inspirehep.net/record/1305430}{Inspire}].

\bibitem{Khachatryan:2014tva}
{\bfseries CMS} Collaboration, V.~Khachatryan {\em et al.,}
  \href{http://dx.doi.org/10.1103/PhysRevD.91.092005}{{\em Phys. Rev.}
  {\bfseries D 91} no.~9, (2015) 092005},
  \href{http://arxiv.org/abs/1408.2745}{{\tt arXiv:1408.2745}}.
  [\href{http://inspirehep.net/record/1310653}{Inspire}].

\bibitem{Khachatryan:2014fba}
{\bfseries CMS} Collaboration, V.~Khachatryan {\em et al.,}
  \href{http://dx.doi.org/10.1007/JHEP04(2015)025}{{\em JHEP} {\bfseries 04}
  (2015) 025}, \href{http://arxiv.org/abs/1412.6302}{{\tt arXiv:1412.6302}}.
  [\href{http://inspirehep.net/record/1335131}{Inspire}].

\bibitem{Rainwater:2007qa}
D.~Rainwater and T.~M.~P. Tait,
  \href{http://dx.doi.org/10.1103/PhysRevD.75.115014}{{\em Phys. Rev.}
  {\bfseries D 75} (2007) 115014},
  \href{http://arxiv.org/abs/hep-ph/0701093}{{\tt hep-ph/0701093}}.
  [\href{http://inspirehep.net/record/742288}{Inspire}].

\bibitem{Alves:2014cda}
D.~S.~M. Alves, J.~Galloway, J.~T. Ruderman and J.~R. Walsh,
  \href{http://dx.doi.org/10.1007/JHEP02(2015)007}{{\em JHEP} {\bfseries 1502}
  (2015) 007}, \href{http://arxiv.org/abs/1410.6810}{{\tt arXiv:1410.6810}}.
  [\href{http://inspirehep.net/record/1324364}{Inspire}].

\bibitem{Brensing:2007qm}
S.~Brensing, S.~Dittmaier, M.~Kramer and A.~Muck,
  \href{http://dx.doi.org/10.1103/PhysRevD.77.073006}{{\em Phys. Rev.}
  {\bfseries D 77} (2008) 073006}, \href{http://arxiv.org/abs/0710.3309}{{\tt
  arXiv:0710.3309}}. [\href{http://inspirehep.net/record/764541}{Inspire}].

\bibitem{Dittmaier:2009cr}
S.~Dittmaier and M.~Huber,
  \href{http://dx.doi.org/10.1007/JHEP01(2010)060}{{\em JHEP} {\bfseries 01}
  (2010) 060}, \href{http://arxiv.org/abs/0911.2329}{{\tt arXiv:0911.2329}}.
  [\href{http://inspirehep.net/record/836737}{Inspire}].

\bibitem{Note1}
These modified propagators encapsulate all new physics effects because they are
  written in the field basis where the vector boson interactions with fermions
  are identical to those of the SM, once expressed in terms of the input
  parameters $\alpha _{\protect \rm {em}}$, $G_F$, and $m_Z$. This explains the
  mismatch with Ref.~\cite {Barbieri:2004qk}, where a different basis is used.

\bibitem{Falkowski:2015krw}
A.~Falkowski and K.~Mimouni,
  \href{http://dx.doi.org/10.1007/JHEP02(2016)086}{{\em JHEP} {\bfseries 1602}
  (2016) 086}, \href{http://arxiv.org/abs/1511.07434}{{\tt arXiv:1511.07434}}.
  [\href{http://inspirehep.net/record/1406141}{Inspire}].

\bibitem{Aaltonen:2012bp}
{\bfseries CDF} Collaboration, T.~Aaltonen {\em et al.,}
  \href{http://dx.doi.org/10.1103/PhysRevLett.108.151803}{{\em Phys. Rev.
  Lett.} {\bfseries 108} (2012) 151803},
\href{http://arxiv.org/abs/1203.0275}{{\tt arXiv:1203.0275}}.

\bibitem{Abazov:2012bv}
{\bfseries D0} Collaboration, V.~M. Abazov {\em et al.,}
  \href{http://dx.doi.org/10.1103/PhysRevLett.108.151804}{{\em Phys. Rev.
  Lett.} {\bfseries 108} (2012) 151804},
\href{http://arxiv.org/abs/1203.0293}{{\tt arXiv:1203.0293}}.

\bibitem{Catani:2007vq}
S.~Catani and M.~Grazzini,
  \href{http://dx.doi.org/10.1103/PhysRevLett.98.222002}{{\em Phys. Rev. Lett.}
  {\bfseries 98} (2007) 222002},
  \href{http://arxiv.org/abs/hep-ph/0703012}{{\tt hep-ph/0703012}}.
  [\href{http://inspirehep.net/record/745577}{Inspire}].

\bibitem{Gavin:2012sy}
R.~Gavin, Y.~Li, F.~Petriello and S.~Quackenbush,
  \href{http://dx.doi.org/10.1016/j.cpc.2012.09.005}{{\em Comput. Phys.
  Commun.} {\bfseries 184} (2013) 208},
  \href{http://arxiv.org/abs/1201.5896}{{\tt arXiv:1201.5896}}.
  [\href{http://inspirehep.net/record/1086536}{Inspire}].

\bibitem{Ball:2012cx}
R.~D. Ball, V.~Bertone, S.~Carrazza, C.~S. Deans, L.~Del~Debbio, S.~Forte,
  A.~Guffanti, N.~P. Hartland, J.~I. Latorre, J.~Rojo and M.~Ubiali, ,
  \href{http://dx.doi.org/10.1016/j.nuclphysb.2012.10.003}{{\em Nucl. Phys.}
  {\bfseries B 867} (2013) 244}, \href{http://arxiv.org/abs/1207.1303}{{\tt
  arXiv:1207.1303}}. [\href{http://inspirehep.net/record/1121392}{Inspire}].

\bibitem{Ball:2013hta}
R.~D. Ball, V.~Bertone, S.~Carrazza, L.~Del~Debbio, S.~Forte, A.~Guffanti,
  N.~P. Hartland and J.~Rojo, ,
  \href{http://dx.doi.org/10.1016/j.nuclphysb.2013.10.010}{{\em Nucl. Phys.}
  {\bfseries B 877} (2013) 290}, \href{http://arxiv.org/abs/1308.0598}{{\tt
  arXiv:1308.0598}}. [\href{http://inspirehep.net/record/1246369}{Inspire}].

\bibitem{Ball:2014uwa}
{\bfseries NNPDF} Collaboration, R.~D. Ball {\em et al.,}
  \href{http://dx.doi.org/10.1007/JHEP04(2015)040}{{\em JHEP} {\bfseries 1504}
  (2015) 040}, \href{http://arxiv.org/abs/1410.8849}{{\tt arXiv:1410.8849}}.
  [\href{http://inspirehep.net/record/1325552}{Inspire}].

\bibitem{Manohar:2016nzj}
A.~Manohar, P.~Nason, G.~P. Salam and G.~Zanderighi,
  \href{http://arxiv.org/abs/1607.04266}{{\tt arXiv:1607.04266}}.
  [\href{http://inspirehep.net/record/1475703}{Inspire}].

\bibitem{ATLAS:2014wra}
{\bfseries ATLAS} Collaboration, G.~Aad {\em et al.,}
  \href{http://dx.doi.org/10.1007/JHEP09(2014)037}{{\em JHEP} {\bfseries 09}
  (2014) 037}, \href{http://arxiv.org/abs/1407.7494}{{\tt arXiv:1407.7494}}.
  [\href{http://inspirehep.net/record/1308524}{Inspire}].

\bibitem{Aaboud:2016zkn}
{\bfseries ATLAS} Collaboration, M.~Aaboud {\em et al.,}
  \href{http://arxiv.org/abs/1606.03977}{{\tt arXiv:1606.03977}}.
  [\href{http://inspirehep.net/record/1469070}{Inspire}].

\bibitem{Baer:2013cma}
H.~Baer {\em et al.,} \href{http://arxiv.org/abs/1306.6352}{{\tt
  arXiv:1306.6352}}. [\href{http://inspirehep.net/record/1240098}{Inspire}].

\bibitem{Gomez-Ceballos:2013zzn}
M.~Bicer {\em et al.,} \href{http://dx.doi.org/10.1007/JHEP01(2014)164}{{\em
  JHEP} {\bfseries 1401} (2014) 164},
  \href{http://arxiv.org/abs/1308.6176}{{\tt arXiv:1308.6176}}.
  [\href{http://inspirehep.net/record/1251418}{Inspire}].

\bibitem{CEPC-SPPCStudyGroup:2015csa}

C.-S.~S. Group.

\bibitem{Fan:2014axa}
J.~Fan, M.~Reece and L.-T. Wang,
  \href{http://dx.doi.org/10.1007/JHEP08(2015)152}{{\em JHEP} {\bfseries 08}
  (2015) 152},
\href{http://arxiv.org/abs/1412.3107}{{\tt arXiv:1412.3107}}.

\bibitem{Harigaya:2015yaa}
K.~Harigaya, K.~Ichikawa, A.~Kundu, S.~Matsumoto and S.~Shirai,
  \href{http://dx.doi.org/10.1007/JHEP09(2015)105}{{\em JHEP} {\bfseries 1509}
  (2015) 105}, \href{http://arxiv.org/abs/1504.03402}{{\tt arXiv:1504.03402}}.
  [\href{http://inspirehep.net/record/1359460}{Inspire}].

\bibitem{Racco:2015dxa}
D.~Racco, A.~Wulzer and F.~Zwirner,
  \href{http://dx.doi.org/10.1007/JHEP05(2015)009}{{\em JHEP} {\bfseries 1505}
  (2015) 009}, \href{http://arxiv.org/abs/1502.04701}{{\tt arXiv:1502.04701}}.
  [\href{http://inspirehep.net/record/1345165}{Inspire}].

\bibitem{Busoni:2013lha}
G.~Busoni, A.~De~Simone, E.~Morgante and A.~Riotto,
  \href{http://dx.doi.org/10.1016/j.physletb.2013.11.069}{{\em Phys. Lett.}
  {\bfseries B 728} (2014) 412}, \href{http://arxiv.org/abs/1307.2253}{{\tt
  arXiv:1307.2253}}. [\href{http://inspirehep.net/record/1242005}{Inspire}].

\bibitem{Contino:2016jqw}
R.~Contino, A.~Falkowski, F.~Goertz, C.~Grojean and F.~Riva,
  \href{http://dx.doi.org/10.1007/JHEP07(2016)144}{{\em JHEP} {\bfseries 1607}
  (2016) 144}, \href{http://arxiv.org/abs/1604.06444}{{\tt arXiv:1604.06444}}.
  [\href{http://inspirehep.net/record/1450010}{Inspire}].

\bibitem{Note2}
This is not completely correct in the charged DY case because low transverse
  mass bins might in principle still receive contributions from reactions that
  occur at very high center of mass energies, well above the cutoff. These
  contribution are however negligible for the analysis discussed in this paper.
  To show this we recalculated the bounds on ${\protect \rm {\protect \sc
  {W}}}$ shown in Table~\ref {tab:boost} artificially including in the
  calculation of the New Physics cross section only those events in which the
  lepton-neutrino invariant mass is below the maximal cutoff $\Lambda
  =m_W/\protect \sqrt {{\protect \rm {\protect \sc {W}}}}$ at which the
  derivative expansion breaks down. These new bounds are only weaker than the
  old ones by a few percent, showing that the contamination mentioned above is
  numerically irrelevant.

\bibitem{Thamm:2015zwa}
A.~Thamm, R.~Torre and A.~Wulzer,
  \href{http://dx.doi.org/10.1007/JHEP07(2015)100}{{\em JHEP} {\bfseries 1507}
  (2015) 100}, \href{http://arxiv.org/abs/1502.01701}{{\tt arXiv:1502.01701}}.
  [\href{http://inspirehep.net/record/1343133}{Inspire}].

\end{thebibliography}\endgroup

\end{document}